\documentclass{iopart}

\usepackage{graphicx}
\usepackage{iopams}
\usepackage{dcolumn}

\newcommand{\ket}[1]{|#1\rangle}

\newcommand{\eps}{\varepsilon}

\begin{document}

\title[Tunnel magnetoresistance of chains of quantum dots]
{The tunnel magnetoresistance in chains of quantum dots weakly
coupled to external leads}

\author{Ireneusz Weymann}
\address{Physics Department, Arnold
Sommerfeld Center for Theoretical Physics, and Center for
NanoScience, Ludwig-Maximilians-Universit\"at, Theresienstrasse
37, 80333 Munich, Germany, and Department of Physics, Adam
Mickiewicz University, Umultowska 85, 61-614 Pozna\'n, Poland}
\ead{weymann@amu.edu.pl}

\date{\today}

\begin{abstract}
We analyze numerically the spin-dependent transport through
coherent chains of three coupled quantum dots weakly connected to
external magnetic leads. In particular, using the diagrammatic
technique on the Keldysh contour, we calculate the conductance,
shot noise and tunnel magnetoresistance (TMR) in the sequential
and cotunneling regimes. We show that transport characteristics
greatly depend on the strength of the interdot Coulomb
correlations, which determines the spacial distribution of
electron wave function in the chain. When the correlations are
relatively strong, depending on the transport regime, we find both
negative TMR as well as TMR enhanced above the Julliere value,
accompanied with negative differential conductance (NDC) and
super-Poissonian shot noise. This nontrivial behavior of tunnel
magnetoresistance is associated with selection rules that govern
tunneling processes and various high-spin states of the chain that
are relevant for transport. For weak interdot correlations, on the
other hand, the TMR is always positive and not larger than the
Julliere TMR, although super-Poissonian shot noise and NDC can
still be observed.
\end{abstract}

\pacs{72.25.Mk, 73.63.Kv, 85.75.-d, 73.23.Hk}

\maketitle

\section{Introduction}

Tunnel magnetoresistance (TMR) is a measure of system transport
properties change when the magnetic configuration of the device
switches from parallel to antiparallel alignment
\cite{julliere75}. The tunneling current is usually larger in the
parallel configuration, when transport occurs between the
majority-majority and minority-minority spin bands, than in the
antiparallel configuration, where electrons tunnel between
majority and minority spin bands, which gives rise to positive TMR
effect. The TMR has been analyzed in various systems, including
single-electron transistors and quantum
dots~\cite{barnas98,bulka00,rudzinski01,koenigPRL03,braunPRB04,
weymannPRB05,braigPRB05,weymannQDs,souzaPRB07,
franssonEPL05,cottetPRB04-06,kondoSQD,sindelPRB07,weymannPRB08,barnasJPCM08}.
In fact, a great deal of theoretical and experimental
investigations has been devoted to spin-polarized transport
through quantum dot structures. This is because quantum dots
coupled to ferromagnetic leads are ideal candidates to study the
fundamental interactions between spins and charges
\cite{wolf01,loss02,maekawa02,zutic04}. Furthermore, such systems
are also being considered for applications in future spintronic
devices as well as for quantum computing \cite{lossPRA98}.
However, most of existing theoretical considerations of
spin-dependent transport in quantum dots involved only single and
double dot systems
\cite{bulka00,rudzinski01,koenigPRL03,braunPRB04,
weymannPRB05,braigPRB05,weymannQDs,souzaPRB07,
franssonEPL05,cottetPRB04-06,kondoSQD,sindelPRB07,weymannPRB08,
barnasJPCM08}, while experiments were carried out mainly for
single dot structures
\cite{zhao02,jensen05,sahoo05,pasupathy04,fertAPL06,hamayaAPL07a,
hamayaAPL07b,hamayaPRB08,parkinNL08,HauptmannNat08}. In
particular, it has been shown~\cite{weymannPRB05} that the TMR in
quantum dots weakly coupled to ferromagnetic leads is generally
smaller than the value given by the Julliere
model~\cite{julliere75}, ${\rm TMR}^{\rm Jull} = 2p^2/(1-p^2)$,
where $p$ is the spin polarization of the leads, which is
characteristic of tunneling through a single tunnel junction. This
result is rather intuitive, as by embedding a quantum dot
structure between ferromagnetic electrodes, the tunneling
processes through the system become incoherent due to spin-flip
processes and spin relaxation in the dot, leading to suppressed
TMR. Because the magnitude of TMR is generally conditioned by the
interplay of spin-dependent tunneling, spin accumulation and
various spin states that mediate the current, one may expect that
in the case of multi-dot structures, where some high-spin
molecular states may form, the behavior of TMR will be modified as
compared to that observed in the case of single and double dots.

To prove the above statement, in this paper we address the problem
of tunneling through chains of quantum dots, consisting of three
coherent dots, weakly coupled to external ferromagnetic leads.
Very recently transport properties of triple quantum dots have
become a subject of intensive studies due to various interesting
effects that emerge in such structures
\cite{LossPRL03,JiangPRB05,GuevaraPRB06,aghassiAPL06,
Kuzmenko06,emaryPRB07,HawrylakPRB07,
TrochaPRB08,Vernek_arXiv09,KostyrkoPRB09,GossardPRL95,
HawrylakPRL06,SchroeerPRB07,GoldmanAPL07,Fujisawa_arXvi08,
RoggePRB08,peresPRB09,kimPRB09,gaudreauPRB09}. In particular,
triple dots enable the investigation of spin-entangled currents
\cite{LossPRL03}, dark states \cite{emaryPRB07,KostyrkoPRB09}, or
various interference effects \cite{GuevaraPRB06,TrochaPRB08}.
Although nonmagnetic properties of multi-dot structures have
already been addressed both theoretically and experimentally, very
little is known about their magnetic transport properties
\cite{TrochaJPCM08}. The goal of this paper is therefore to
discuss the spin-polarized transport through coherent triple
quantum dots. In particular, by employing the real-time
diagrammatic technique, we calculate the current, differential
conductance, TMR and shot noise in both the sequential and
cotunneling regimes. We show that transport characteristics
strongly depend on the strength of the interdot correlations,
which determines the spacial distribution of electron wave
functions in the chain. In the case of strong Coulomb
correlations, we find that the TMR may take values larger than the
Julliere TMR, which is associated with tunneling through high-spin
molecular states of the quantum dot system. Moreover, we also
predict negative TMR, due to an increased tunneling current in the
antiparallel configuration, associated with spin accumulation in
the chain. In addition, we show that these effects may be
accompanied with super-Poissonian shot noise and negative
differential conductance (NDC). On the other hand, in the case of
weak interdot Coulomb correlations, the TMR is always positive and
not larger than the Julliere TMR, while we still observe
super-Poissonian shot noise in the Coulomb blockade regime and
negative differential conductance. Although here we consider
chains consisting of only three quantum dots, similar behavior may
be also observed in longer chains where transport occurs through
high-spin molecular states and is governed by various selection
rules.

The paper is organized as follows. In the second section we
describe the Hamiltonian of the quantum dot chain and briefly
discuss the method used in calculations. Section III is devoted to
numerical results, where we first analyze the transport
characteristics in the case of strong interdot Coulomb
correlations and then proceed to discuss the transport behavior in
the case of weak interdot correlations. Finally, the conclusions
are given in Sec.~IV.

\section{Theoretical description}

\subsection{Model}

\begin{figure}[t]
  \includegraphics[width=0.7\columnwidth]{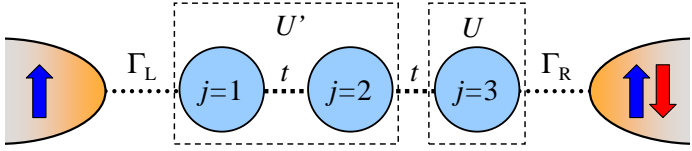}
  \caption{\label{Fig:1} (color online)
  The schematic of chain consisting of three
  single-level quantum dots connected to external
  ferromagnetic leads. The hopping between neighboring dots
  is denoted by $t$, $U'$ and $U$ are the
  inter-dot and intra-dot Coulomb correlation energies,
  while $\Gamma_{\rm L}$ and $\Gamma_{\rm R}$
  denote the couplings to the left and right leads.
  The magnetizations of the leads are assumed to
  form either parallel or antiparallel magnetic configuration,
  as sketched in the figure.
  }
\end{figure}

The schematic of chain consisting of three quantum dots coupled to
ferromagnetic leads is shown in Fig.~\ref{Fig:1}. It is assumed
that the magnetizations of the leads are oriented collinearly, so
that the system can be either in the parallel or antiparallel
magnetic configuration. The Hamiltonian of the system is given by
\begin{equation}\label{Eq:H}
   H = H_{\rm lead} + H_{\rm tun} + H_{\rm chain} \,,
\end{equation}
where the first part corresponds to noninteracting itinerant
electrons in the left ($r={\rm L}$) and right ($r={\rm R}$) lead,
$H_{\rm lead} = \sum_r \sum_{{\mathbf k}\sigma}
\varepsilon_{r{\mathbf k}\sigma} c^{\dagger}_{r{\mathbf k}\sigma}
c_{r{\mathbf k}\sigma}$, where $\varepsilon_{r{\mathbf k}\sigma}$
is the energy of an electron with the wave vector ${\mathbf k}$
and spin $\sigma$ in the lead $r$, and $c^{\dagger}_{r{\mathbf
k}\sigma}$ ($c_{r{\mathbf k}\sigma}$) denotes the respective
creation (annihilation) operator. The second term of
Eq.~(\ref{Eq:H}) accounts for the tunneling processes between the
leads and the quantum dot chain,
\begin{equation}
  H_{\rm tun}=\sum_{\mathbf k\sigma} \left(t_{\rm
  L}c^{\dagger}_{{\rm L} {\mathbf k}\sigma} d_{1\sigma} + t_{\rm
  R}c^{\dagger}_{{\rm R} {\mathbf k}\sigma} d_{3\sigma} + {\rm h.c.}
  \right) \;,
\end{equation}
where $t_{r}$ denotes the tunnel matrix elements between the lead
$r$ and the respective dot and $d_{j\sigma}$ destroys a
spin-$\sigma$ electron in the dot $j$ ($j=1,2,3$). Note, that the
first dot is coupled to the left lead, while the third dot is
connected to the right lead, see Fig.~\ref{Fig:1}. The strength of
the coupling of the quantum dot chain to the spin-majority
(spin-minority) electron band of the $r$th lead is given by,
$\Gamma_r ^ {+(-)} = 2\pi |t_r|^2\rho_r^{+(-)} = \Gamma_r (1\pm
p_r)$, where $\Gamma_r= (\Gamma_r^{+} +\Gamma_r^{-})/2$, while
$\rho_r^{+(-)}$ and $p_r$ are the spin-dependent density of states
for majority (minority)  spin band
and spin polarization in the lead $r$, respectively. In the
following we assume $\Gamma_{\rm L} = \Gamma_{\rm R} \equiv
\Gamma/2$ and $p_{\rm L} = p_{\rm R} \equiv p$.

Finally, the last term of the Hamiltonian describes the chain
consisting of three quantum dots which is given by
\begin{eqnarray}\label{Eq:Hchain}
  H_{\rm chain} &=&
  \sum_{j} \sum_{\sigma} \varepsilon_{j} n_{j\sigma}
  + \sum_{j} U_{j} n_{j\uparrow}n_{j\downarrow}
  \nonumber\\
  &+& U^\prime \sum_{\sigma\sigma^\prime}
  \left(n_{1\sigma}n_{2\sigma^\prime} + n_{2\sigma}n_{3\sigma^\prime}\right)
     \nonumber\\
  &+&
  t \sum_{\sigma} \left(
  d_{1\sigma}^\dagger d_{2\sigma}+d_{2\sigma}^\dagger d_{3\sigma} +h.c. \right)
 \,,
\end{eqnarray}
with $n_{j\sigma}=d^{\dagger}_{j\sigma}d_{j\sigma}$ begin the
particle number operator on dot $j$, while $\varepsilon_{j}$ and
$U_j$ denote the single-particle energy and on-level Coulomb
correlation in dot $j$, respectively. The third part of $H_{\rm
chain}$ corresponds to the inter-dot Coulomb interaction, whose
strength is given by $U^\prime$, while $t$ describes the inter-dot
hopping. As we are interested in rather low bias voltage regime,
it is justifiable to assume that the dot energy levels are
independent of the bias voltage. This assumption has also been
verified numerically and even if one assumes a voltage drop on the
outer dots of the order of $10\%$, the current flowing through the
system becomes only very slightly modified as compared that in the
case of zero voltage drop. Therefore, for the sake of clarity of
further discussion, we take the energy-independent dot levels.
Furthermore, we also assume that the system is symmetric, i.e.,
$\varepsilon_j \equiv \varepsilon$ and $U_j \equiv U$ ($j=1,2,3$).

\subsection{Method}

In order to calculate the spin-polarized transport through a chain
of three coherent quantum dots in the sequential and cotunneling
regimes, we employ the real-time diagrammatic technique
\cite{weymannPRB05,diagrams,thielmannPRL05}. It generally consists
in a perturbative expansion of the density matrix of the system
and the operators of interest (for example the current operator)
with respect to the coupling strength $\Gamma$. Time evolution of
the reduced density matrix is given by a sequence of irreducible
self-energy blocks, $\Sigma_{\chi \chi^\prime}$, on the Keldysh
contour, corresponding to various transition events between the
many-body states $\ket{\chi}$ and $\ket{\chi^\prime}$ of the
quantum dot chain. On the other hand, the full propagation of the
reduced density matrix is given by the Dyson equation, which can
be further transformed into a general kinetic equation. In a
steady state the kinetic equation is simply given by
$(\mathbf{\tilde{\Sigma}}\mathbf{P})_{\chi} =
\Gamma\delta_{\chi\chi_0}$\,,
and enables the calculation of occupation probabilities $P_\chi$
for the system to be in a many-body state $\ket{\chi}$. Here,
$\mathbf{P}$ is the probability vector, while the matrix
$\mathbf{\tilde{\Sigma}}$ is given by the self-energy matrix
$\mathbf{\Sigma}$ with one arbitrary row $\chi_0$ replaced by
$(\Gamma,\dots,\Gamma)$ due to normalization,
$\sum_{\chi}P_{\chi}=1$. The current flowing through the system
can be then found from \cite{diagrams}
\begin{equation}\label{Eq:current}
  I=-\frac{ie}{2\hbar}{\rm Tr}\{\mathbf{\Sigma}^{\rm I}\mathbf{P}\} \,,
\end{equation}
where $\mathbf{\Sigma}^{\rm I}$ denotes the modified self-energy
matrix $\mathbf{\Sigma}$ so as to take into account the number of
electrons transferred through the system.

To calculate the transport properties order by order in tunneling
processes, we expand the self-energy matrices, $\mathbf{\Sigma} =
\mathbf{\Sigma}^{(1)} +\mathbf{\Sigma}^{(2)}+\dots$,
$\mathbf{\Sigma}^{\rm I} = \mathbf{\Sigma}^{\rm I(1)}
+\mathbf{\Sigma}^{\rm I(2)}+\dots$, and the occupations,
$\mathbf{P} = \mathbf{P}^{(0)} + \mathbf{P}^{(1)}+\dots$,
respectively. The self-energies in respective order can be then
calculated using the corresponding diagrammatic rules
\cite{weymannPRB05,diagrams}. The first order of expansion
corresponds to the sequential tunneling, whereas the second one to
cotunneling. In this analysis we have calculated the self-energies
up to the second order of perturbation series, so that we are able
to resolve transport properties both in the sequential and
cotunneling regimes \cite{weymannPRB08}. The sequential tunneling
dominates transport above a threshold voltage and is exponentially
suppressed in the Coulomb blockade regime \cite{nato92}. In the
blockade regime, on the other hand, the dominant contribution to
the current comes from cotunneling processes \cite{cotunneling},
which take place through virtual states of the system and are only
algebraically suppressed in the Coulomb blockade. As the influence
of cotunneling on transport for bias voltages above the threshold
for sequential tunneling is rather minor, the inclusion of
second-order processes is crucial for a proper description of
transport in the blockade regimes.

In addition, in the following we will also analyze the
zero-frequency current noise \cite{blanterPR00},
$S=\int_{-\infty}^\infty dt (\langle
\hat{I}(t)\hat{I}(0)+\hat{I}(0)\hat{I}(t)\rangle-2 \langle
\hat{I}\rangle^2 )$, where $\hat{I}$ is the current operator,
$\hat{I} = (\hat{I}_{\rm R}-\hat{I}_{\rm L})/2$, with
$\hat{I}_{\rm L(R)} = -i(e/\hbar) t_{\rm L (R) } \sum_{{\mathbf
k}\sigma} (c^{\dagger}_{\rm L(R) {\mathbf k}\sigma}
d_{1(3)\sigma}- d^\dagger_{1(3)\sigma} c_{\rm L(R) {\mathbf
k}\sigma})$ being the current flowing from the first (third) dot
to the left (right) lead. The formula for current noise derived
within the real-time diagrammatic technique can be found in
Ref.~\cite{thielmannPRL05}.

\section{Numerical results}

In the following we will discuss the numerical results on the
current, differential conductance, tunnel magnetoresistance and
the shot noise of a chain of tree coherent single-level quantum
dots in both the linear and nonlinear response regimes. Transport
characteristics of such systems strongly depend on the internal
parameters, in particular, on the ratio between inter-dot Coulomb
repulsion $U'$ and the hopping between the dots $t$, provided that
$U>U',|t|$. The ratio can be tuned experimentally for example by
changing the height of the barrier between the dots
\cite{Simmons_arXiv09}. When the inter-dot Coulomb correlations
are relatively strong, $U'/|t|>1$, the electrons in the ground
state of the chain will be mostly occupying the outermost dots. On
the other hand, for weak inter-dot Coulomb interactions,
$U'/|t|<1$, this tendency will not be observed. Thus, depending on
$U'/|t|$, the spacial distribution of the many-body chain states
may become strongly modified. In this paper we will therefore
discuss the transport characteristics in the two above mentioned
situations. Furthermore, we also note that due to many intrinsic
parameters of the system, there is a variety of transport regimes
where different transport behavior can be observed. In the
following, we will thus present general density plots, however
only most interesting transport features will be discussed in
greater detail.

\begin{figure*}[t]
  \includegraphics[width=1\columnwidth]{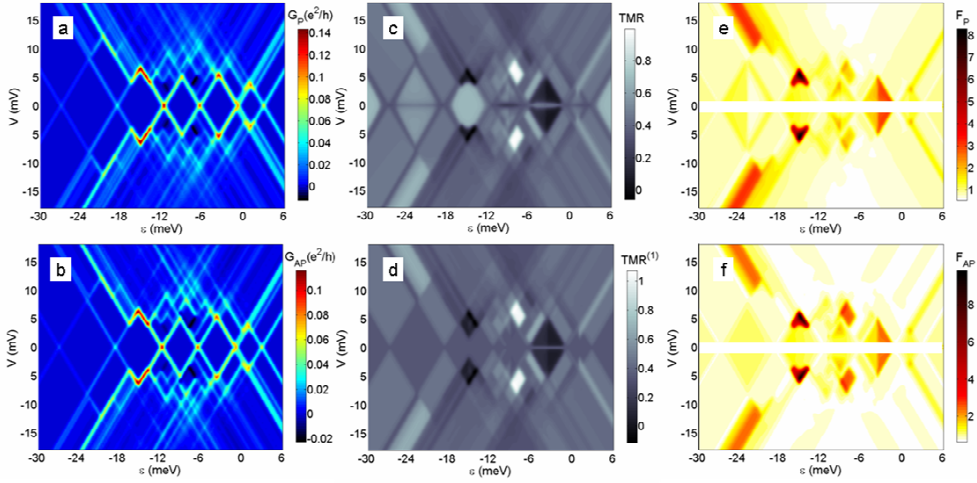}
  \caption{\label{Fig:2} (color online)
  The total (first plus second order) differential conductance in the parallel (a)
  and antiparallel (b) alignment, the total TMR (c),
  the TMR calculated using only first-order processes (d)
  and the total Fano factor for parallel (e) and antiparallel (f)
  configuration of the system.
  The figures were calculated for the case where the inter-dot
  Coulomb interaction is larger than the inter-dot hopping, $U'/|t|>1$.
  The parameters are: $U^\prime =
  4$ meV, $t=-2$ meV, $U=10$ meV, $k_BT=0.15$ meV,
  $\Gamma=0.1$ meV and $p=0.5$.
  The Fano factor at low bias voltages diverges,
  therefore this transport regime is
  marked with white stripe in (e) and (f).}
\end{figure*}

\subsection{The case of strong inter-dot Coulomb interactions}

Figure~\ref{Fig:2} shows various transport characteristics of a
quantum dot chain as a function of bias voltage $V$ and the
position of the dots' levels $\varepsilon$. Because the position
of the levels can be experimentally changed by sweeping the gate
voltage, Fig.~\ref{Fig:2} effectively presents a bias and gate
voltage dependence of transport characteristics. The total (first
plus second order) differential conductance for the parallel
($G_{\rm P}$) and antiparallel ($G_{\rm AP}$) magnetic
configurations is shown in Fig.~\ref{Fig:2}(a) and (b),
respectively. First of all, one can see that the differential
conductance displays characteristic Coulomb diamond pattern, with
Coulomb blockade regimes at low transport voltages. By lowering
the position of the dots' levels the chain becomes consecutively
occupied with electrons. In the case considered here, the quantum
dot chain can accommodate up to six electrons, i.e. each dot can
be doubly occupied. For such values of $\varepsilon$ when the two
neighboring charge states become degenerate, there is a peak in
the linear conductance. On the other hand, with increasing the
bias voltage, out of the Coulomb blockade regime, there are
additional lines visible in the differential conductance
associated with tunneling through excited states of the system.
Furthermore, due to the contact with ferromagnetic leads, the
tunneling processes in the system become spin-dependent and,
consequently, transport depends on the magnetic configuration of
the system. In the parallel configuration the majority (minority)
electrons of one lead tunnel to the majority (minority) spin band
of the other lead, whereas in the antiparallel configuration they
tunnel to the minority (majority) spin band. This is why the
conductance in the antiparallel configuration is generally
suppressed as compared to that in the parallel configuration,
$G_{\rm P}>G_{\rm AP}$, see Fig.~\ref{Fig:2}(a) and (b). This
difference in turn gives rise to nonzero tunnel magnetoresistance
which is plotted in Fig.~\ref{Fig:2}(c).

The TMR reflects the change of system transport properties when
switching the magnetic configuration of the device from parallel
to antiparallel one. It is qualitatively defined
as~\cite{julliere75,barnas98,rudzinski01}, ${\rm TMR}=I_{\rm
P}/I_{\rm AP}-1$, where $I_{\rm P}$ ($I_{\rm AP}$) it the current
flowing through the system in the parallel (antiparallel) magnetic
configuration. Usually the conductance in the parallel
configuration is larger than that in the antiparallel one, giving
rise to positive TMR. In particular, for a single ferromagnetic
tunnel junction the TMR can be described by the Julliere model
\cite{julliere75}, ${\rm TMR^{Jull}}=2p^2/(1-p^2)$ (${\rm
TMR^{Jull}}=2/3$ for $p=0.5$ assumed in calculations).
Intuitively, one may expect that placing a quantum dot molecule
between the two leads (where tunneling processes are generally
incoherent) will decrease the TMR. This is in fact what is
observed in most quantum dot structures -- for symmetric systems
and in the absence of magnetic field, the TMR in the weak coupling
regime is positive and not larger than ${\rm TMR^{Jull}}$
\cite{weymannPRB05,barnasJPCM08}. In the case of tunneling through
quantum dot chains considered here, however, we predict a
nontrivial behavior of the TMR. Depending on the transport regime,
we find both the TMR enhanced above the Julliere value as well as
negative TMR effect, see Fig.~\ref{Fig:2}(c). The mechanisms
responsible for these effects will be discussed in more detail in
the sequel.

We also note that the TMR is directly related to the ratio between
the currents in the two magnetic configurations, so that its
magnitude does not necessarily dependent on the magnitude of
tunneling current. This makes the TMR a vary sensitive quantity
for analyzing transport properties in various regimes, especially
where sequential tunneling is suppressed due to Coulomb
correlations and transport occurs mainly through higher-order
tunneling processes. For comparison, in Fig.~\ref{Fig:2}(d) we
have also plotted the TMR calculated using only the first-order
tunneling processes, ${\rm TMR^{(1)}}$. It can be seen that cotunneling processes
modify the TMR mainly in the blockade regimes and give rise to
strong dependence of TMR on the occupation number of the quantum
dot chain. On the other hand, out of the blockade regimes, the current
is predominantly mediated by sequential tunneling and one finds
that ${\rm TMR}$ and ${\rm TMR^{(1)}}$ become comparable,
although not equal.

In addition, we have also calculated the Fano factor, $F=S/S_p$,
in both magnetic configurations, see Fig.~\ref{Fig:2}(e) and (f).
The Fano factor describes the deviation of the shot noise from its
Poissonian value, $S_p=2e|I|$, which is characteristic of
uncorrelated tunneling. When transport is mediated only by elastic
cotunneling processes, the noise is Poissonian, $F\to 1$, however,
once the spin-flip cotunneling is allowed the noise can be
enhanced to become super-Poissonian, $F>1$, due to bunching of
inelastic processes~\cite{sukhorukovPRB01}. Furthermore, in the
sequential tunneling regime, transport is mainly dominated by
Coulomb correlations which decrease the noise and the Fano factor
is generally sub-Poissonian, $F<1$ \cite{thielmannPRB03}. This
behavior can be in fact observed in Fig.~\ref{Fig:2}(e) and (f),
where in the cotunneling regime the Fano factor can take large
super-Poissonian values, while in the sequential tunneling regime
it becomes rather suppressed. It can be also seen that the general
behavior of the Fano factor in the parallel ($F_{\rm P}$) and
antiparallel ($F_{\rm AP}$) magnetic configurations is quite
similar, although the magnitude of the noise is larger in the
antiparallel configuration. On the other hand, in the low bias
voltage regime, the noise is dominated by thermal noise while the
current tends to zero, which leads to a divergency in the Fano
factor. Therefore this transport regime is marked with white
stripes in Fig.~\ref{Fig:2}(e) and (f).

\subsubsection{Linear response regime}

\begin{figure}[t]
  \includegraphics[width=0.5\columnwidth]{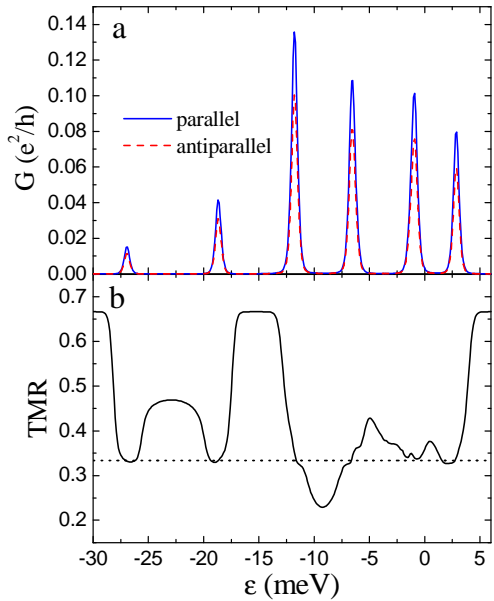}
  \caption{\label{Fig:3} (color online)
  The linear conductance (a) in the parallel (solid line)
  and antiparallel (dashed line) magnetic configurations
  and linear response TMR (b)
  as a function of the dots' level position $\varepsilon$.
  The parameters are the same as in Fig.~\ref{Fig:2}.
  The dotted line in (b) shows the linear TMR calculated
  by using only first-order processes -- sequential TMR
  is constant and given by ${\rm TMR^{(1)}}=p^2/(1-p^2)$.}
\end{figure}

In the linear response regime, the transport behavior is mainly
conditioned by the ground state of the system and its evolution
when changing the position of the dots levels. The ground state
energies $E^{\rm G}_{Q,S}$ together with respective quantum
numbers of states $\{Q,S\}$ are shown in Table~\ref{tab:energies},
with
\begin{eqnarray}
   Q=\sum_{j\sigma}n_{j\sigma}\;,
   \hspace{0.5cm}
   \vec{S}=\frac{1}{2}\sum_{j\sigma\sigma'}d_{j\sigma}^\dagger
   \vec{\sigma}_{\sigma\sigma'}d_{j\sigma'}\;,
\end{eqnarray}
denoting the total charge and total spin of the quantum dot chain.
Because in the absence of external magnetic field the Hamiltonian
of the chain, Eq.~(\ref{Eq:Hchain}), commutes with $Q$ and $S^2$,
one can solve the eigenvalue problem by diagonalizing $H_{\rm
chain}$ in respective blocks $\{Q,S\}$. Furthermore, by using the
full spin $SU(2)$ symmetry the size of the Hilbert space is effectively reduced
from $64$ to $35$ multiplets, which may be crucial for analytical discussion
of decoupled quantum dot chain. In numerical calculations,
however, we have used the $64$-state space with the following
many-body states: $\ket{\chi}=\ket{\chi_1\chi_2\chi_3}$, where
$\chi_j=0,\uparrow,\downarrow,{\rm d}$ denotes the state with zero
electrons, one spin-up, spin-down electron and two electrons on
the dot $j$. This is because in the case of the spin-dependent
coupling to ferromagnetic leads, the Hamiltonian of the whole
system possesses only the $S_z$ symmetry. In
Table~\ref{tab:energies} we show the corresponding quantum
numbers, the dimension of the Hamiltonian blocks, degeneracy of
states and ground state energies of the decoupled quantum dot
chain. It turns out that the ground state energies $E^{\rm
G}_{Q,S}$ for smaller blocks can be easily calculated, however,
for larger blocks the formulas become too lengthy to be presented
here. We thus list the explicit expressions for low-dimension
Hamiltonian blocks, while for the other blocks we just state which
energy is the lowest one in respective charge sector $Q$. The explicit
matrices for Hamiltonian blocks $H_{Q,S}$ together with the
definition of states for total charge and total spin symmetries
can be found in the Appendix.

\begin{table}[t]
\caption{\label{tab:energies} The charge $Q$ and the total spin
$S$ of three quantum dots coupled in series, dimension of
respective $\{Q,S\}$ blocks, $D$, degeneracy of states, $d$, and
ground state energies, $E^{\rm G}_{Q,S}$. The analytical formulas
for ground state energies of larger blocks are rather lengthy,
therefore here we only state which states have lower energy for
given $Q$. The explicit matrices for Hamiltonian blocks are given
in the Appendix.} \centering
\begin{tabular}{llllll}
$n$\;\;\;\;& $Q$ & $S$\;\;\; & $D$ & $d$\;\;\;\; & Ground states energies\\
\hline
1 & $0$ & $0$ & $1$ & $1$ & $E^{\rm G}_{0,0}=0$\vspace{0.12cm}\\
2 & $1$ & $\frac{1}{2}$ & $3$ & $2$ & $E^{\rm G}_{1,\frac{1}{2}}=\varepsilon-\sqrt{2}|t|$\vspace{0.12cm}\\
3 & $2$ & $0$ & $6$ & $1$ & $E^{\rm G}_{2,0}\lesssim E^{\rm G}_{2,1}$\vspace{0.12cm}\\
4 & $2$ & $1$ & $3$ & $3$ & $E^{\rm G}_{2,1}=2\varepsilon + \frac{1}{2}U' - \sqrt{2t^2+(U'/2)^2}$\vspace{0.12cm}\\
5 & $3$ & $\frac{1}{2}$ & $8$ & $2$ & $E^{\rm G}_{3,\frac{1}{2}} \lesssim E^{\rm G}_{3,\frac{3}{2}}$\vspace{0.12cm}\\
6 & $3$ & $\frac{3}{2}$ & $1$ & $4$ & $E^{\rm G}_{3,\frac{3}{2}}=3\varepsilon+2U'$\vspace{0.12cm}\\
7 & $4$ & $0$ & $6$ & $1$ & $E^{\rm G}_{4,0}\lesssim E^{\rm G}_{4,1}$\vspace{0.12cm}\\
8 & $4$ & $1$ & $3$ & $3$ & $E^{\rm G}_{4,1}=4\varepsilon+U+\frac{7}{2}U'-\sqrt{2t^2+(U'/2)^2}$\vspace{0.12cm}\\
9 & $5$ & $\frac{1}{2}$ & $3$ & $2$ & $E^{\rm G}_{5,\frac{1}{2}}=5\varepsilon+2U+5U'-\sqrt{2t^2+U'^2}$\vspace{0.12cm}\\
10 & $6$ & $0$ & $1$ & $1$ & $E^{\rm G}_{6,0}=6\varepsilon+3U+8U'$
\end{tabular}
\end{table}

The linear conductance as well as the total TMR are shown in
Fig.~\ref{Fig:3}(a) and (b). The linear conductance displays
characteristic resonance peaks whenever two neighboring charge
states become degenerate. The resonance energies can be estimated
from Table~\ref{tab:energies} by solving ${\rm min}\{ E^{\rm
G}_{Q+1,S}\}-{\rm min}\{E^{\rm G}_{Q,S}\}=0$, where one needs to
take the minimum energy for given $Q$. The conductance in the
parallel configuration is larger than the conductance in the
antiparallel configuration, see Fig.~\ref{Fig:3}(a), which results
in positive linear TMR, see Fig.~\ref{Fig:3}(b). For comparison we
have also plotted the TMR obtained using only first-order
tunneling processes, which is constant and given by ${\rm
TMR^{(1)}}=p^2/(1-p^2)$, see the dotted line in
Fig.~\ref{Fig:3}(b). The total TMR, on the other hand, shows a
nontrivial dependence on the position of the dots' levels
$\varepsilon$. As shown in the case of single quantum dots
\cite{weymannPRB05,barnasJPCM08}, the magnitude of linear TMR is
directly related to the type of cotunneling processes that drive
the current in respective transport regimes. Among various
cotunneling events, one can distinguish processes that affect the
magnetic state of the quantum dot system (inelastic spin-flip
processes) and the ones that do not affect the quantum dot chain
(elastic non-spin-flip processes). In the case when each dot of
the chain is either empty or doubly occupied, only elastic
processes are possible, however, in the other cases, the spin-flip
processes become also allowed.

When the chain is empty ($Q=0$) or fully occupied ($Q=6$), TMR is
maximal and equal to the Julliere value, ${\rm TMR}=2p^2/(1-p^2)$,
see Fig.~\ref{Fig:3}(b). This is related to the fact that in these
transport regimes only the non-spin-flip cotunneling processes are
allowed and transport is fully coherent -- the co-tunneling
electrons are not scattered at the chain at all. Interestingly,
also for $Q=4$ the TMR becomes equal to ${\rm TMR^{Jull}}$, which
indicates that only elastic cotunneling contributes to the linear
conductance. In fact, for $Q=4$ the ground state of the chain is
$S=0$, see Table~\ref{tab:energies}. Furthermore, for assumed
parameters, i.e. in the case of strong interdot correlations,
$U'/t>1$, it turns out that in the ground state the chain is
occupied with two electrons in the outermost dots, so that the ground
state is $\ket{\rm d0d}$. In this case only the non-spin-flip
cotunneling is allowed, which yields the maximum TMR. However, the
situation changes once the spin-flip processes become possible,
which happens in the other transport regimes. In particular, for
$Q=1,2,3$, the linear TMR becomes suppressed to approximately a
half of ${\rm TMR^{Jull}}$ and its dependence on $\varepsilon$ is
rather complex. In these regimes the current is mainly mediated by
inelastic spin-flip cotunneling. For $Q=5$, on the other hand, the
TMR becomes slightly enhanced, although it is still lower than
${\rm TMR^{Jull}}$. Because for $Q=5$ the ground state is a
doublet $\ket{\rm d\sigma d}$, transport is due to both elastic
and inelastic processes -- the former (latter) ones tend to
increase (decrease) the TMR, so that the magnitude of TMR is
between the values corresponding to $Q=1,2,3$ and $Q=0,4,6$
transport regimes. Finally, we also note that at resonances the
total TMR drops to the value approximately given by ${\rm
TMR}^{(1)}$, as for resonant energies the first-order processes
become possible and are dominant.

\subsubsection{Enhanced TMR and negative differential conductance}

\begin{figure}[t]
  \includegraphics[width=0.5\columnwidth]{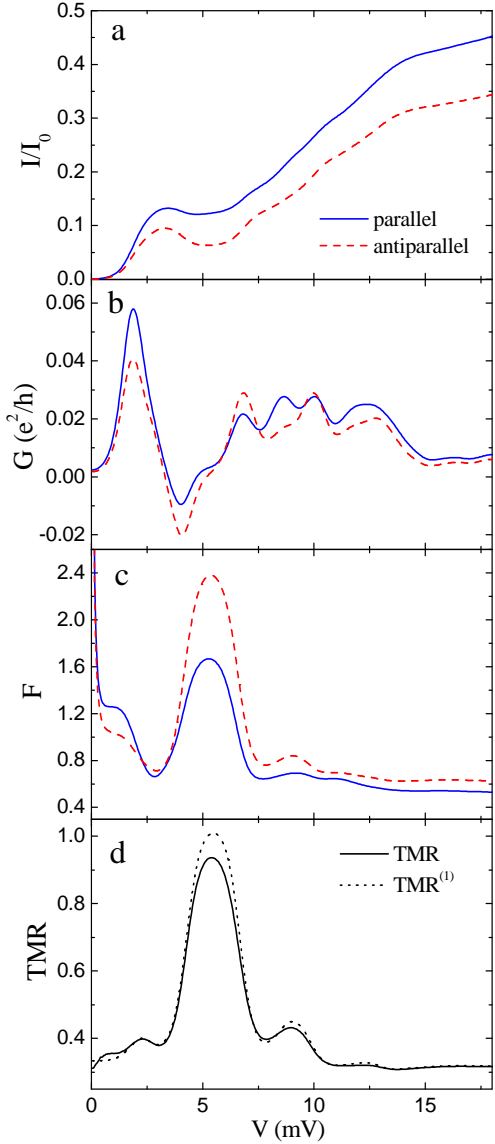}
  \caption{\label{Fig:4} (color online)
  The bias voltage dependence of the current (a),
  differential conductance (b),
  Fano factor (c) in the parallel (solid line)
  and antiparallel (dashed line) magnetic configurations
  and the resulting TMR (d) for $\varepsilon=-7.5$ meV.
  The other parameters are the same as in Fig.~\ref{Fig:2} and
  $I_0$ denotes the maximum current given by $I_0=e\Gamma/\hbar$.}
\end{figure}

The bias dependence of the current, differential conductance, Fano
factor and TMR is shown in Fig.~\ref{Fig:4}. The transport
characteristics were calculated for $\varepsilon=-7.5$ meV, see
also Fig.~\ref{Fig:2}, which corresponds to the case when at
equilibrium the quantum dot chain is in the spin doublet state,
$\{Q=3,S=\frac{1}{2}\}$, so that the ground state is doubly
degenerate. With increasing the bias voltage, more and more states
start participating in transport and the current increases.
However, it can be seen that the bias dependence of the current is
not monotonic -- after the first Coulomb step, the current starts
to decrease with $V$, leading to negative differential conductance
(NDC), which is present in both magnetic configurations, see
Fig.~\ref{Fig:4}(b). The suppression of the current is associated
with selection rules that govern the respective sequential
transitions, i.e. only the transitions that change the total
charge of the chain by $1$ and the total spin by $\frac{1}{2}$ are
allowed. When raising the transport voltage, the following excited
states $\{Q,S\}$ become active in transport: $\{2,0\}$ singlet and
$\{2,1\}$ triplet, then $\{3,\frac{1}{2}\}$ doublet, and then
$\{3,\frac{3}{2}\}$ quadruplet, respectively. Although the
sequential transitions between $Q=3$ doublet and $Q=2$ singlet and
triplet are possible, the transitions between $Q=3$ doublet and
$Q=3$ quadruplet are prohibited as they obey neither the charge
nor the spin selection rules. In fact, once the system gets
trapped in the quadruplet state, transport becomes suppressed, see
Fig.~\ref{Fig:4}(a) for $V\approx 5$ mV, which leads to negative
differential conductance. This is because transitions involving
$\{3,\frac{3}{2}\}$ state can occur only through tunneling
from/into the triplet state $\{2,1\}$. In the case of $U/|t'|>1$,
the two electrons in the triplet state are localized in the
outermost dots, while in the quadruplet state the three electrons
are distributed uniformly between the dots. Thus, for a transition
between $\{3,\frac{3}{2}\}$ and $\{2,1\}$ to occur, one needs to
put or remove an electron from the middle dot, which is however
suppressed as $(t/U')^2$. On the other hand, tunneling processes
in which the state of the system changes between the $Q=3$ doublet
and $Q=2$ singlet are rather independent of the ratio $t/U'$, as
they can occur through the outmost dots. Consequently, transitions
involving the quadruplet state are relatively slow, while the
other ones are much faster. The competition between such {\it
slow} and {\it fast} transport channels may in turn lead to large
current fluctuations \cite{aghassiAPL06}. This can be seen in
Fig.~\ref{Fig:4}(c), where for voltages corresponding to the
transport regime where the current is suppressed, super-Poissonian
shot noise is observed. When increasing the bias voltage further,
more excited states become available for transport and the current
starts increasing again, see Fig.~\ref{Fig:4}(a), while the shot
noise becomes suppressed to sub-Poissonian value, which is typical
of charge-correlated sequential transport, see
Fig.~\ref{Fig:4}(c).

In the antiparallel configuration, on the other hand, due to the
asymmetry of tunneling processes between the left and right leads,
there is a nonequilibrium spin accumulation in the chain. For
positive bias voltages, the occupation probability of
highest-weight spin states is much increased as compared to the
other components of particular state. This is because the spin-up
electrons tunneling from the left lead to the chain and the
spin-down electrons tunneling out of the chain to the right lead
belong to the majority-spin bands, and the positive spin component
becomes accumulated in the chain. Consequently, fewer states are
available for transport as compared to the parallel configuration,
so that in the antiparallel configuration the transport channel
involving the quadruplet state $\{3,\frac{3}{2}\}$ becomes even
less transmitting. This leads to several interesting features.
First of all, the current is more suppressed in the antiparallel
configuration than in the parallel one which leads to an enhanced
NDC, see Fig.~\ref{Fig:4}(b). On the other hand, this more
effective current suppression also reflects itself in an enhanced
super-Poissonian shot noise, see Fig.~\ref{Fig:4}(c). Furthermore,
for voltages where the current is suppressed we observe the TMR
enhanced above the Julliere value. The TMR in this transport
regime is approximately given by ${\rm TMR}\approx \frac{3}{2}
{\rm TMR^{Jull}}$, see Fig.~\ref{Fig:4}(d). As shown in previous
considerations, such enhancement of TMR in serial quantum dots
weakly coupled to external leads can occur mainly in asymmetric
systems or in the presence of external magnetic field
\cite{weymannPRB08,barnasJPCM08}. Here, we observe ${\rm
TMR>TMR^{Jull}}$ in the absence of magnetic field and for fully
symmetric system. This is just a generic feature of transport
through chains of quantum dots, where due to selection rules the
system may be trapped in some high-spin states.

\subsubsection{Negative TMR and super-Poissonian shot noise}

\begin{figure}[t]
  \includegraphics[width=0.5\columnwidth]{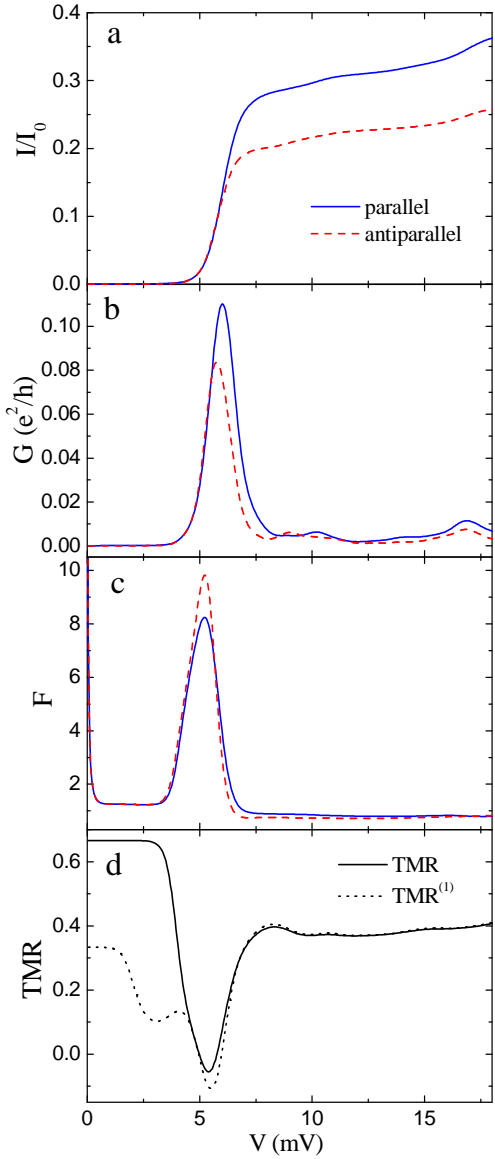}
  \caption{\label{Fig:5} (color online)
  The bias voltage dependence of the current (a),
  differential conductance (b),
  Fano factor (c) in the parallel (solid line)
  and antiparallel (dashed line) magnetic configurations
  and the resulting TMR (d) for $\varepsilon=-15$ meV
  and the parameters as in Fig.~\ref{Fig:2}.}
\end{figure}

Another interesting transport behavior can be observed in the case
where the chain is in the ground state with four electrons. The
current, differential conductance, shot noise and TMR as a
function of the bias voltage for $\eps=-15$ meV are shown in
Fig.~\ref{Fig:5}. Due to large interdot Coulomb correlations,
$U'/|t|>1$, the ground state is non-degenerate, with doubly
occupied outermost dots, $\ket{\rm d0d}$. The nearest excited
states are respectively: doublets $\{3,\frac{1}{2}\}$ and
$\{5,\frac{1}{2}\}$, and triplet $\{4,1\}$. These states are
relatively close to each other and start taking part in transport
for voltages around the threshold for sequential tunneling. At low
bias voltages the system is in the Coulomb blockade, see
Fig.~\ref{Fig:5}(a) and (b), and transport is due to elastic
cotunneling processes, which yield the Poissonian shot noise and
TMR given by the Julliere value, see Fig.~\ref{Fig:5}(c) and (d).
The situation changes once the transport voltage approaches the
threshold, $V\approx 5$ mV, then the TMR suddenly drops and
changes sign, while the shot noise becomes strongly enhanced. This
is associated with tunneling processes that become allowed in this
transport regime. First of all, the inelastic cotunneling
processes become possible for each doublet state, i.e.
$\{3,\frac{1}{2}\}$ and $\{5,\frac{1}{2}\}$. Furthermore, around
the threshold voltage, the sequential processes also start
participating in transport. The first-order transitions occur
first between the $Q=4$ singlet and $Q=3$, $Q=5$ doublets. It is
worth noting that the spacial distribution of the wave function is
different for these two doublets. For $Q=3$ the electrons are
equally distributed between the three dots, while for $Q=5$ the
outermost dots are fully occupied while the middle dot is singly
occupied to minimize the interdot correlations. Consequently, the
transport channel involving the state $\{5,\frac{1}{2}\}$ is
slower than that involving the state $\{3,\frac{1}{2}\}$,
similarly as in the case discussed in previous subsection. It
turns out that the interplay of various first and second-order
tunneling processes, where particular events occur at different
rates, which exists for transport voltages around the threshold
for sequential tunneling, leads to large current fluctuations. As
a result, we observe an enhanced super-Poissonian shot noise in
both magnetic configurations of the system, see
Fig.~\ref{Fig:5}(c). On the other hand, when the voltage increases
further, sequential processes dominate transport and the noise
becomes generally sub-Poissonian.

An interesting transport feature visible around the threshold
voltage is the negative TMR effect, see Fig.~\ref{Fig:5}(d). To
understand this behavior, one needs to realize a very delicate
difference between probability distributions in the two magnetic
configurations. In the Coulomb blockade regime the chain is in the
singlet state $\ket{\rm d0d}$ with probability equal to unity,
irrespective of magnetic configuration of the system. However,
once the bias voltage approaches the threshold voltage, the
occupation probability of states $\{3,\frac{1}{2}\}$,
$\{5,\frac{1}{2}\}$, and $\{4,1\}$ starts slowly increasing. In
addition, it turns out that in the antiparallel configuration the
probability of the highest-weight quadruplet state is also
nonzero, and it is slightly larger than the occupation
probabilities of the above-mentioned doublets and triplet. The
enhanced occupation probability of $\{3,\frac{3}{2}\}$ is purely
due to nonequilibrium spin accumulation, it is thus not present in
the parallel configuration. This is in fact what is crucial for
the occurrence of negative TMR. In the antiparallel configuration
the current can in addition flow due to cotunneling and
thermally-activated first-order transitions involving the
quadruplet state, which is not possible in the parallel
configuration. As a result, the current in the antiparallel
configuration becomes larger than the current in the parallel
configuration, yielding negative TMR effect. This can be also seen
in the differential conductance where the first peak in the
antiparallel configuration occurs at slightly lower bias voltage
than in the parallel one, see Fig.~\ref{Fig:5}(b). With increasing
the bias voltage further, the excited states start participating
in transport and the system apparently exhibits a {\it normal}
spin-valve behavior \cite{braunPRB04,weymannPRB05}, with the
current in the parallel configuration larger than in the
antiparallel one and, thus, with positive TMR, see
Fig.~\ref{Fig:5}(d). It is interesting to note that the increased
occupation probability of the highest-spin component of the
quadruplet state in the antiparallel configuration was also
responsible for the enhanced TMR effect discussed in previous
subsection, whereas here it lead to negative TMR. The negative TMR
however occurs on the edge of the Coulomb blockade regime, while
the enhanced TMR develops in the sequential tunneling regime.

\subsection{The case of weak inter-dot Coulomb interaction}

\begin{figure}[t]
  \includegraphics[width=\columnwidth]{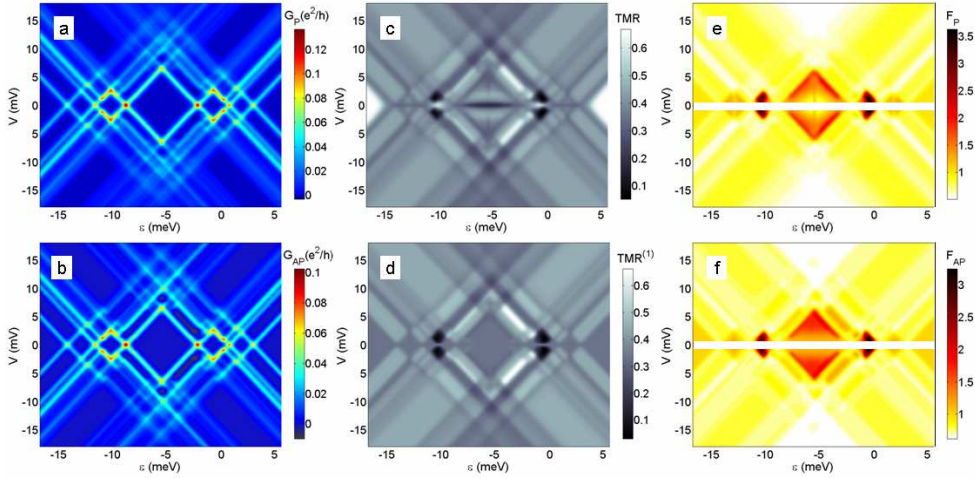}
  \caption{\label{Fig:6} (color online)
  The total differential conductance in the parallel (a)
  and antiparallel (b) alignment, the total TMR (c),
  the TMR calculated using only first-order processes (d)
  and the total Fano factor for parallel (e) and antiparallel (f)
  configuration of the system
  for the case when the inter-dot
  Coulomb interaction is smaller than the inter-dot hopping, $U'/|t|<1$.
  The parameters are: $U^\prime =
  0.5$ meV, $t=-2$ meV, $U=10$ meV, $k_BT=0.15$ meV,
  $\Gamma=0.1$ meV and $p=0.5$.
  The Fano factor at low bias voltage diverges,
  therefore this transport regime is
  marked with white stripe in (e) and (f).}
\end{figure}

The differential conductance, TMR and Fano factor in the case of
$U'/|t|<1$ are shown in Fig.~\ref{Fig:6}. For $U'/|t|<1$, the
electrons in particular states are distributed rather uniformly
over the three dots, contrary to the previous case where electrons
where localized in the outermost dots to minimize the Coulomb
correlation energy. This results in a more symmetric behavior of
transport characteristics with respect to the middle of the
Coulomb blockade regime with $Q=3$, which is due to particle-hole
symmetry. Moreover, as most of the effects observed $U'/|t|>1$
were mainly associated with spacial distribution of the wave
function, one may expect their strong dependence on the ratio
$U'/|t|$. This is in fact what can be observed. For example, when
$U'/|t|<1$, the negative TMR and TMR enhanced above the Julliere
value are not present, although super-Poissonian shot noise and
negative differential conductance can still be found in some
transport regimes. Of course, the difference between transport
characteristics in the case of weak and strong interdot
correlations reveals itself mainly in transport regimes where the
states with more than a single electron become relevant, see
Figs.~\ref{Fig:2} and \ref{Fig:6}.

The differential conductance in the parallel and antiparallel
configurations is shown in Fig.~\ref{Fig:6}(a) and (b),
respectively. First of all, we note that because the energy of
interdot correlations is now changed, the Coulomb diamond
structure is different from that shown in Fig.~\ref{Fig:2}. The
largest Coulomb diamond develops for $Q=3$, while the size of
other diamonds is much decreased, except for empty and fully
occupied chain, see Fig.~\ref{Fig:6}(a) and (b). The conductance
in the parallel configuration is larger than in the antiparallel
one and the TMR is positive in the whole range of bias voltage $V$
and the level position $\varepsilon$, see Fig.~\ref{Fig:6}(c). For
comparison, the TMR calculated using only the sequential tunneling
processes is shown in Fig.~\ref{Fig:6}(d). The main difference
between the density plots for the total ${\rm TMR}$ and ${\rm
TMR^{(1)}}$ can be seen in the Coulomb blockade regimes where
cotunneling dominates the current. It can be seen that the total
TMR in the linear response regime displays nontrivial dependence
on the occupation number of the chain. For empty and fully
occupied chain, TMR is given by the Julliere value and it is much
suppressed in other blockade regimes due to spin-flip cotunneling
processes.

In the nonlinear response regime when at equilibrium the chain was
in the charge state $Q=3$, there is reminiscent of effects found
in the case of $U'/|t|>1$, see Fig.~\ref{Fig:4}. Now, one can also
observe an enhanced TMR, although its magnitude is slightly lower
than the Julliere TMR. This enhanced TMR is accompanied with
negative differential conductance and super-Poissonian shot noise,
which are more visible in the antiparallel configuration. The
mechanism leading to these effects is similar to that discussed in
the previous subsection and is mainly associated with transport
channel involving the quadruplet state and spin accumulation in
the antiparallel configuration.

Furthermore, in the nonlinear response regime of the Coulomb
blockade regime with $Q=2$, the TMR becomes suppressed, being very
close to zero. In this transport regime the chain at equilibrium
is in the singlet state $\{2,0\}$, and the excited states are
consecutively $\{2,1\}$ and $\{1,\frac{1}{2}\}$. With increasing
the bias voltage, occupation probability of excited states starts
increasing. It turns out that in the parallel configuration all
the components of $Q=1$ doublet and $Q=2$ triplet are relevant,
while in the antiparallel configuration, due to spin accumulation,
only the highest-weight components, however with slightly larger
occupation probabilities. This leads to an increased current in
the antiparallel configuration, so that the currents in the two
configurations become roughly comparable, yielding very small TMR.
In addition, with raising the bias voltage the shot noise becomes
enhanced and reaches maximum for voltages around the threshold for
sequential tunneling, which is associated with bunching of
inelastic cotunneling processes. Similar behavior can be also
observed for the Coulomb blockade regime with four electrons in
the chain, which is due to particle-hole symmetry. It is also
interesting to note that the behavior of transport characteristics
in the cotunneling regime with $Q=2$ is very weakly affected by
the ratio $U'/|t|$, see Figs.~\ref{Fig:2} and \ref{Fig:6}.
However, for Coulomb blockade regimes with more electrons in the
ground state, transport properties become completely modified due
to different spacial distribution of wave functions, compare for
example the Coulomb blockade regime with $Q=4$ in
Figs.~\ref{Fig:2} and \ref{Fig:6}.

Finally, we also note that in the case of weak interdot
correlations the shot noise is rather sub-Poissonian in the whole
sequential tunneling regime, irrespective of magnetic
configuration of the system. On the other hand, the
super-Poissonian shot noise is only found in the Coulomb blockade
regimes where the chain is the charge state with two, three or
four electrons, which is due to bunching of inelastic cotunneling
processes.

\section{Conclusions}

We have analyzed the linear and nonlinear transport properties of
chains of quantum dots consisting of three coherent single-level
quantum dots weakly coupled to external ferromagnetic leads. By
employing the real-time diagrammatic technique, we have calculated
the current, differential conductance, shot noise and tunnel
magnetoresistance in the case of strong ($U'/|t|>1$) and weak
($U'/|t|<1$) interdot correlations. By changing the ratio
$U'/|t|$, one can effectively change the spacial distribution of
electron wave functions of the chain. When $U'/|t|>1$, the
electrons tend to be localized in the outermost dots, while for
$U'/|t|<1$ the electrons are distributed rather uniformly over the
dots.

In particular, in the case of large interdot correlations we have
shown that the TMR strongly depends on the transport regime and
can take negative values as well as values exceeding the TMR given
by the Julliere model. The enhanced TMR occurs in the nonlinear
response regime when the chain is occupied by three electrons at
equilibrium and is associated with a suppressed current in the
antiparallel configuration due to trapping of the quantum dot
chain in some high-spin states. In addition, the suppression of
the current gives rise to negative differential conductance and
super-Poissonian shot noise. We have also shown that the TMR may
change sign and become negative. This happens in the cotunneling
regime with four electrons in the chain when the bias voltage
approaches threshold voltage for sequential tunneling. The
negative TMR is then associated with increased tunneling through
the highest-weight spin state (quadruplet) of the chain.
Furthermore, we have also shown that negative TMR is accompanied
with large super-Poissonian shot noise due to the interplay
between various inelastic cotunneling and sequential processes
that start contributing to the current around the threshold
voltage. On the other hand, when the interdot correlations are
weak, most of the effects found in the case of $U'/|t|>1$ become
smeared out. In particular, the negative TMR and TMR enhanced
above the Julliere value are not present, although
super-Poissonian shot noise and the negative differential
conductance can still be observed.

Finally, we note that although the results presented here were
calculated for chains of three quantum dots, similar behavior may
be in principle observed for longer chains, where transport is
governed by various selection rules and the current can flow due
to tunneling through high-spin molecular states of the system.


\ack

We acknowledge discussions with J. Barna\'s. This work was
supported by funds of the Polish Ministry of Science and Higher
Education as a research project for years 2008-2010, by the
Alexander von Humboldt Foundation and the Foundation for Polish
Science. Financial support by the Excellence Cluster "Nanosystems
Initiative Munich (NIM)" is gratefully acknowledged.


\appendix

\section*{Appendix}

The localized basis of quantum dot chain states is defined as
$\ket{\chi}=\ket{\chi_1\chi_2\chi_3}$, where
$\chi_j=0,\uparrow,\downarrow,{\rm d}$ denotes zero electrons, one
spin-up, spin-down electron and doubly occupied dot $j$. Using the
$SU(2)$ symmetry for total spin one can reduce the Hilbert space
from $4^3$ states to $35$ multiplets. The reduction of Hilbert space is
important rather for analytical calculations, while in
numerical calculations we have used the basis of $64$ states. In
the following, we give the explicit matrices for blocks of the
Hamiltonian $H_{\rm chain}$ in the basis of total charge $Q$ and
total spin $S$, where to define the spin $SU(2)$ basis we have taken
the highest-weight spin states. In the block $\{Q=2,S=0\}$, the
states are: $S_1^{2,0}=\ket{{\rm d}00}$, $S_2^{2,0}=\ket{\rm
0d0}$, $S_3^{2,0}=\ket{\rm 00d}$,
$S_{12}^{2,0}=\frac{1}{\sqrt{2}}(\ket{\uparrow\downarrow
0}-\ket{\downarrow\uparrow 0})$,
$S_{23}^{2,0}=\frac{1}{\sqrt{2}}(\ket{0\uparrow\downarrow
}-\ket{0\downarrow\uparrow})$, and $S_{13}^{2,0} =
\frac{1}{\sqrt{2}}(\ket{\uparrow 0\downarrow}-\ket{\downarrow
0\uparrow})$, respectively. The Hamiltonian block in this basis is
given by
\begin{equation}
  H_{2,0} = \left(
\begin{array}{cccccc}
  \eps_{21} & 0 & 0 & \sqrt{2}t & 0 & 0\\
  0 & \eps_{21} & 0 & \sqrt{2}t & \sqrt{2}t & 0\\
  0 & 0 & \eps_{21} & 0 & \sqrt{2}t & 0\\
  \sqrt{2}t & \sqrt{2}t & 0 & \eps_{22} & 0 & t\\
  0 & \sqrt{2}t & \sqrt{2}t & 0 & \eps_{22} & t\\
  0 & 0 & 0 & t & t & \eps_{23}
\end{array}
\right), \end{equation}
where $\eps_{21}=2\eps+U$, $\eps_{22}=2\eps+U'$, and
$\eps_{23}=2\eps$. The block of $H_{\rm chain}$ for $\{Q=4,S=0\}$,
$H_{4,0}$, has similar structure to $H_{2,0}$ due to particle hole
symmetry,
\begin{equation}
  H_{4,0} = \left(
\begin{array}{cccccc}
  \eps_{41} & 0 & 0 & \sqrt{2}t & 0 & 0\\
  0 & \eps_{41}-4U' & 0 & \sqrt{2}t & \sqrt{2}t & 0\\
  0 & 0 & \eps_{41} & 0 & \sqrt{2}t & 0\\
  \sqrt{2}t & \sqrt{2}t & 0 & \eps_{42} & 0 & -t\\
  0 & \sqrt{2}t & \sqrt{2}t & 0 & \eps_{42} & -t\\
  0 & 0 & 0 & -t & -t & \eps_{43}
\end{array}
\right), \end{equation}
where $\eps_{41}=4\eps+2U+4U'$, $\eps_{42}=4\eps+U+3U'$, and
$\eps_{43}=4\eps+U+4U'$. The states in block $\{Q=4,S=0\}$ are
explicitly given by: $S_1^{4,0}=\ket{0\rm dd}$,
$S_2^{4,0}=\ket{\rm d0d}$, $S_3^{4,0}=\ket{\rm dd0}$,
$S_{12}^{4,0}=\frac{1}{\sqrt{2}}(\ket{\uparrow\downarrow \rm d} -
\ket{\downarrow\uparrow \rm d})$, $S_{23}^{4,0} =
\frac{1}{\sqrt{2}}(\ket{\rm d\uparrow\downarrow } - \ket{\rm d
\downarrow\uparrow})$, and $S_{13}^{4,0} =
\frac{1}{\sqrt{2}}(\ket{\uparrow \rm d\downarrow}-\ket{\downarrow
\rm d\uparrow})$, respectively. On the other hand, the Hamiltonian
block for $\{Q=3,S=\frac{1}{2}\}$ is the largest one, with the
states defined as follows: $D_1^{3,\frac{1}{2}}=\ket{\rm d\uparrow
0}$, $D_2^{3,\frac{1}{2}} = \ket{\rm 0\uparrow d}$,
$D_3^{3,\frac{1}{2}} = \ket{\rm \uparrow d 0}$,
$D_4^{3,\frac{1}{2}} = \ket{\rm 0d\uparrow}$, $D_5^{3,\frac{1}{2}}
= \ket{\rm \uparrow 0 d}$, $D_6^{3,\frac{1}{2}} = \ket{\rm d
0\uparrow}$, $D_7^{3,\frac{1}{2}} = \frac{1}{\sqrt{2}}
(\ket{\uparrow\downarrow\uparrow}-
\ket{\downarrow\uparrow\uparrow})$, and $D_8^{3,\frac{1}{2}} =
\sqrt{\frac{2}{3}}\ket{\uparrow\uparrow\downarrow}-\frac{1}{\sqrt{6}}
(\ket{\uparrow\downarrow\uparrow}+\ket{\downarrow\uparrow\uparrow})$,
respectively. The block $H_{3,\frac{1}{2}}$ is explicitly given by
\begin{equation}
  H_{3,\frac{1}{2}} = \left(
\begin{array}{cccccccc}
  \eps_{31} & 0 & -t & 0 & 0 & t & 0 & 0\\
  0 & \eps_{31} & 0 & -t & t & 0 & 0 & 0\\
  -t & 0 & \eps_{31} & 0 & 0 & 0 & \frac{-t}{\sqrt{2}} & \sqrt{\frac{3}{2}}t\\
  0 & -t & 0 & \eps_{31} & 0 & 0 & \sqrt{2}t & 0\\
  0 & t & 0 & 0 & \eps_{32} & 0 & \frac{-t}{\sqrt{2}} & \sqrt{\frac{3}{2}}t\\
  t & 0 & 0 & 0 & 0 & \eps_{32} & \sqrt{2}t & 0\\
  0 & 0 & \frac{-t}{\sqrt{2}} & \sqrt{2}t & \frac{-t}{\sqrt{2}} & \sqrt{2}t & \eps_{33} & 0\\
  0 & 0 & \sqrt{\frac{3}{2}}t & 0 & \sqrt{\frac{3}{2}}t & 0 & 0 & \eps_{33}
\end{array}
\right), \end{equation}
where $\eps_{31}=3\eps+U+2U'$, $\eps_{32}=3\eps+U$, and
$\eps_{33}=3\eps+2U'$. For completeness, we also give the matrices
for smaller blocks of the Hamiltonian,
\begin{eqnarray}
  H_{1,\frac{1}{2}} &=& \left(
\begin{array}{ccc}
  \eps & t & 0\\
  t & \eps & t\\
  0 & t & \eps
  \end{array}
\right) ,\\
  H_{2,1} &=& \left(
\begin{array}{ccc}
  2\eps+U' & t & 0\\
  t & 2\eps & t\\
  0 & t & 2\eps+U'
  \end{array}
\right).
\end{eqnarray}
The states for block $\{Q=1,S=\frac{1}{2}\}$ are:
$D_{1}^{1,\frac{1}{2}}=\ket{\uparrow 00}$,
$D_{2}^{1,\frac{1}{2}}=\ket{0\uparrow 0}$,
$D_{3}^{1,\frac{1}{2}}=\ket{00\uparrow }$, while for block
$\{Q=2,S=1\}$ they are given by:
$T_{1}^{1,1}=\ket{\uparrow\uparrow 0}$, $T_{2}^{1,1}=\ket{\uparrow
0\uparrow}$, $T_{3}^{1,1}=\ket{0\uparrow\uparrow }$, respectively.
The Hamiltonian blocks $H_{5,\frac{1}{2}}$ and $H_{4,1}$ have
similar structure to blocks $H_{1,\frac{1}{2}}$ and $H_{2,1}$ due
to particle hole symmetry. On the other hand, blocks $H_{0,0}$,
$H_{3,\frac{3}{2}}$ and $H_{6,0}$ are trivially one dimensional,
see also Table~\ref{tab:energies}.


\end{document}